\documentclass[preprint]{agujournal2019}
\usepackage{natbib}
\usepackage{hyperref}
\usepackage{amsmath}
\usepackage{amssymb}
\usepackage{url} 
\usepackage{hyperref}
\usepackage[inline]{trackchanges} 
\usepackage{soul}
\draftfalse

\journalname{Earth and Space Science}

\begin{document}

%
%


\title{Multitaper magnitude-squared coherence for time series with missing data: Understanding oscillatory processes traced by multiple observables}

%
%




\authors{Sarah E. Dodson-Robinson\affil{1}  and Charlotte Haley\affil{2}}

\affiliation{1}{University of Delaware \\
Bartol Research Institute \\
207 Sharp Lab \\
Newark, DE 19716, USA}

\bigskip

\affiliation{2}{Argonne National Laboratory \\
Mathematics and Computer Science Division \\
9700 S Cass Ave \\
Lemont, IL 60439, USA}





\correspondingauthor{Sarah E. Dodson-Robinson}{sdr@udel.edu}



\begin{keypoints}
\item Haley (2021) extended the multitaper method for estimating magnitude-squared coherence (MSC) to accommodate time series with missing data
\item We demonstrate missing-data multitaper MSC by showing that solar Lyman~$\alpha$ flux and Dst index jointly trace known solar midterm oscillations
\item We suggest that multitaper should be the preferred frequency domain method for heliospheric applications due to its superior performance
\end{keypoints}

%
%

%
%


\begin{abstract}

To explore the hypothesis of a common source of variability in two time series, observers may estimate the magnitude-squared coherence (MSC), which is a frequency-domain view of the cross correlation. For time series that do not have uniform observing cadence, MSC can be estimated using Welch's overlapping segment averaging. However, multitaper has superior statistical properties to Welch's method in terms of the tradeoff between bias, variance, and bandwidth. The classical multitaper technique has recently been extended to accommodate time series with underlying uniform observing cadence from which some observations are missing. This situation is common for solar and geomagnetic datasets, which may have gaps due to breaks in satellite coverage, instrument downtime, or poor observing conditions. We demonstrate the scientific use of missing-data multitaper magnitude-squared coherence by detecting known solar mid-term oscillations in simultaneous, missing-data time series of solar Lyman $\alpha$ flux and geomagnetic Disturbance Storm Time index. Due to their superior statistical properties, we recommend that multitaper methods be used for all heliospheric time series with underlying uniform observing cadence.

\end{abstract}

\section*{Plain Language Summary}

The magnitude-squared coherence (MSC) statistic detects oscillations with the same origin that show up in different types of measurements. For example, daily measurements of the air temperature at noon and the number of visitors to an outdoor swimming pool would likely have high MSC at a period of one year. This paper demonstrates how to apply a recently developed multitaper technique for estimating MSC between series of measurements that have some missing values. Our demonstration series are (1) the sun's brightness at a particular wavelength of ultraviolet light and (2) the disturbance of an equatorial ring of current in Earth's magnetosphere. MSC estimates show that both series trace solar oscillations with periods between 50 days and 5 years. Because of its superior statistical properties, we recommend that the multitaper method demonstrated here be widely adopted for analysis of heliospheric measurement series.

%
%

%


%
%
%
%

\section{Introduction} \label{sec:intro}

To maximize understanding of a physical process and build a model with predictive power, an observer will wish to study the process using as many metrics as possible. 
For example, solar physicists probe the sun's magnetic dynamo by combining data from activity-sensitive absorption lines \citep{snow14, machol19}, radio flux indices \citep{tapping13, dudokdewit17}, and vector magnetograms \citep{leka96, sun12}, among many others. The cross-correlation \citep{shumwaystoffer} is a popular tool for examining the manifestation of a single oscillatory process in multiple observables and searching for lead-lag relationships. The fact that solar coronal rotation is modulated by the Hale cycle was revealed by the discovery of a 22-year periodicity in the cross-correlation between yearly averages of the sidereal coronal rotation period and the sunspot number \citep{chandra11}.





However, the functional form of the cross-correlation may be complex if two time series jointly trace multiple oscillations---which will generally be true for heliospheric data since the sun has approximately 10 million weak-amplitude normal modes \citep{pieee07}, one third of which are active at any given time \citep{stix2012sun}. By examining the magnitude-squared coherence (MSC), the observer can distinguish different oscillatory processes, evaluate their statistical significance, and uncover frequency-dependent time lags. In addition, moving into the frequency domain is ideal because many solar and heliospheric time series have underlying uniform observing cadence but are missing some data due to poor observing conditions or instrument downtime, a situation referred to in the literature as the missing-data problem \citep{stoica2000gapes, larsson01, wang2005spectral}. The Sacramento Peak Ca II K-line monitoring program is a case in point: although the goal was to take data every day near noon, observers were only able to do so on 10\% of the days before 1986 and 45\% of the days post-1986 because of weather interruptions and telescope scheduling \citep{white98}. Estimating the cross-correlation between two time series with missing data requires interpolation \citep[e.g.][]{white94, peterson98, mchardy04} or more advanced methodologies such as the z-transformed discrete correlation function \citep{alexander13}. But it is possible to circumvent the interpolation step and compute the MSC using a non-uniform fast Fourier transform (NUFFT) algorithm \citep{keiner2009using}.


Observers wishing to estimate MSC from either interpolated or missing-data time series may choose from three procedures: periodogram smoothing \citep{shumwaystoffer}, overlapping segment averaging \citep{welch67}, and multitapering \citep{thomson1982spectrum, thomson91}. 
But while Welch's method has the advantage of being suitable for time series with irregular observing cadences, multitaper has better statistical properties. For a high-quality power spectrum or coherence estimate, the observer should try to minimize three metrics: (1) bias, (2) variance, and (3) bandwidth, which sets the resolution of the estimator. If one holds two of the three metrics constant, multitaper produces a lower value of the third than the periodogram or Welch's method in the absence of a priori information \citep{bronez92}. 

Although the classical multitaper method of \citet{thomson1982spectrum} can only be applied to time series with uniform observing cadence, two major extensions to statistical theory have made it possible to compute multitaper MSC for time series with missing data. First, \citet{chave2019multitaper} found that the calculations of \citet{bronez1988spectral} could be simplified to produce frequency-independent tapers (to be defined in \S \ref{sec:multitaper}) for the special case of missing data. Next, \citet{H2021} used the missing-data tapers identified in \cite{chave2019multitaper} to define multitaper MSC and phase spectra, which reveal frequency-dependent phase shifts between the two time series. \citet{missingdata} presented a detailed introduction to multitaper power spectrum estimation for astronomical missing-data time series. 

Here we show how to apply the multitaper magnitude-squared coherence method of \cite{H2021} to heliospheric time series with missing data. In \S \ref{sec:explaincoherence} we define the MSC statistic and explain its relationship to the cross-correlation. In \S \ref{sec:multitaper} we review the classical multitaper method \citep{thomson1982spectrum} and then describe the \citet{bronez1988spectral} and \citet{chave2019multitaper} extension for time series with missing data. We explain how to use the missing-data coherence estimator introduced by \cite{H2021} in \S \ref{sec:jackcoh}. To demonstrate the method, we show that solar Lyman $\alpha$ flux and geomagnetic Disturbance Storm Time index jointly trace three well-known midterm quasiperiodic processes (\S \ref{sec:samplecalc}). We present our conclusions in \S \ref{sec:conclusions}.


Missing data bivariate time series are ubiquitous in the physical sciences, engineering, and medicine, and the technique described is invaluable for exploratory analysis without first interpolating gaps. For this reason, the missing data multitaper coherence can be used as a general purpose tool for investigating time series with gaps. This paper introduces the missing data multitaper coherence to the solar physics community on two well-known and well-investigated time series.

\section{Background}
\label{sec:explaincoherence}

Consider two jointly stationary real-valued zero-mean processes $x(t)$ and $y(t)$ with \emph{cross spectrum}
\begin{equation} \label{eq:crosspectrum}
    S_{xy}(f) =  \int_{-\infty}^{\infty} \mbox{E}\{x(t) y(t +\tau)\} e^{-i 2 \pi f \tau} d\tau
\end{equation}
where statistical expectation is denoted by E$\{\cdot\}$ and the cross covariance, $\mbox{E}\{x(t) y(t +\tau)\}$ depends only on the lag $\tau$ due to joint stationarity of the two processes. The \emph{spectrum} of the process $x(t)$ is simply $S_{xx}(f)$. A simple measure for the presence of a common oscillation in $x(t)$ and $y(t)$ at frequency $f$ is the \emph{magnitude squared coherence (MSC)}, denoted $C^2_{xy}(f)$.
\begin{equation}
    C^2_{xy}(f) = \frac{|S_{xy}(f)|^2}{S_{xx}(f) \, S_{yy}(f)}.
    \label{eq:MSCdef}
\end{equation}
 The normalization by $S_{xx}(f) \, S_{yy}(f)$ restricts the MSC to the range $0 \leq C^2_{xy}(f) \leq 1$. MSC can be interpreted as a frequency-dependent correlation coefficient indicating the proportion of variance in common to $x(t)$ and $y(t)$ via a linear relationship.
That is, if one time series can be expressed as a linear combination of lagged versions of the other time series, then a simple linear relationship exists between the two series and the MSC is unity at all frequencies.
 Its associated {\it phase spectrum},
\begin{equation}
    \phi_{xy}(f) = \arctan \left( \frac{\mathfrak{Im} [S_{xy}(f)]}{\mathfrak{Re}[S_{xy}(f)]} \right)
    \label{eq:phasespec}
\end{equation}
(where $\mathfrak{Im}[\cdot]$ and $\mathfrak{Re}[\cdot]$ denote imaginary and real parts, respectively) gives the phase lag as a function of frequency for shared oscillations. 


Suppose $x(t)$ and $y(t)$ are sampled with discrete, unit timesteps, yielding time series $x_n$ and $y_n$,
where $n = 0, 1, 2, \ldots, N-1$, $N$ is the number of observations, and the time step is one unit, $\Delta t = 1$. 
Assume that the sample mean of the observations $\bar{x} = N^{-1} \sum_{n=0}^{N-1} x_n$ is zero, or simply let $x_n$ denote the original series minus its sample mean. 
Define the complex-valued discrete Fourier transform $\widetilde{a}(f)$ of an arbitrary deterministic sequence $a_n$
\begin{equation}
    \widetilde{a}(f) = \sum_{n=0}^{N-1} a_n e^{-2 \pi i f n} .
\end{equation}
Using $\widehat{\cdot}$ to denote a statistical estimator, we write the estimator for the cross spectrum, \eqref{eq:crosspectrum}, as
\begin{equation}
\widehat{S}_{xy}(f) = \widetilde{x}(f) \widetilde{y}^*(f).
    \label{eq:Ecrossspectrum}
\end{equation}

Defining the estimator $\widehat{C}^2_{xy}(f)$ is not as straightforward as substituting the respective spectrum and cross spectrum estimators, as this results in
$\widehat{C}^2_{xy}(f) = \widetilde{x}(f)^2 \, \widetilde{y}^*(f)^2 / \widetilde{x}(f)^2 \, \widetilde{y}(f)^2 \equiv 1$. But assuming the observer can construct $K$ statistically independent estimators $\widehat{S}^{(k)}_{xy}(f)$, $\widehat{S}^{(k)}_{xx}(f)$, and $\widehat{S}^{(k)}_{yy}(f)$, where $k = 0, 1, 2, \ldots, K-1$, the {\it simplest} nontrivial estimator for MSC is
\begin{equation}
    \widehat{C}^2_{xy}(f) = \frac{ \left| \sum_{k=0}^{K-1} \widehat{S}^{(k)}_{xy}(f) \right|^2 } {\left( \sum_{k=0}^{K-1} \widehat{S}^{(k)}_{xx}(f) \right) \; \left( \sum_{k=0}^{K-1} \widehat{S}^{(k)}_{yy}(f) \right)}
    \label{eq:cohest}
\end{equation}
\citep{carter1987coherence}. There are two main ways to create $K$ estimates for the cross spectra: Welch's method and multitaper or a combination of both. In the \citet{welch67} method, the $K$ estimates come from breaking the time series into overlapping segments. Multitaper leverages $K$ orthogonal windows or {\it tapers} to create approximately independent estimates for the cross-spectra, which we will see in the next section.

The distribution of the magnitude squared coherence estimate is given in \citet[Eqn. 2.53]{tc91} or \citet[p 259]{hannan-mts}, given the value for the true coherence. Henceforth we will assume that $C^2_{xy}(f)$ is zero at all frequencies, that is, there is a null hypothesis of no linear relationship between the two random processes. Under this assumption, the distribution is simply
\begin{equation}\label{eq:mscdist}
p(\widehat{C}^2) = \frac{(1-\widehat{C}^2)^{K-2}}{K-1},
\end{equation}
where $p$ denotes probability density.

With a non-uniform fast Fourier transform algorithm, it is possible to estimate MSC for  time series with missing data without interpolating, even though the direct cross-correlation cannot be calculated. Of particular interest here is the ability to estimate MSC using the multitaper method.

\section{Multitaper estimation with missing data}
\label{sec:multitaper}

Applying Equation \eqref{eq:cohest} requires $K$ estimates of three quantities: the power spectrum of $x_n$, the power spectrum of $y_n$, and their cross-spectrum. Denote by $w^{(k)}_n$ a set of $K$ data tapers of length $N$. (This section contains a  brief discussion of how to choose $K$; see \citet[\S 2.7]{missingdata} for a more thorough explanation.) By pointwise multiplying the time series by each $w^{(k)}_n$ before taking a Fourier transform, the observer obtains $K$ sets of {\it eigencoefficients}, denoted $\widetilde{x}^{(k)}(f)$,  
\begin{equation}
    \widetilde{x}^{(k)}(f) = \mathcal{F} \{ w^{(k)}_n x_n \} = \sum_{n=0}^{N-1} w^{(k)}_n (x_n - \bar{x}) e^{-2 \pi i f n} 
    \label{eq:taperedFFT}
\end{equation}
and $\widetilde{y}^{(k)}(f)$, defined similarly. If $x_n$ has missing values---i.e.\ the time series has duration $M$ time units and the $N$ observations occur at integer times $n = 0, \ldots, M$, where $N < M$---the eigencoefficients are
\begin{equation}
    \widetilde{x}^{(k)}(f) = \mathcal{F}^{NU} \{ w^{(k)}_n x_n \},
    \label{eq:taperedNUFFT}
\end{equation}
where $\mathcal{F}^{NU}$ is the NUFFT employed with the $-1$ flag for the exponent \citep{keiner2009using}. 
Then one obtains $k$ estimates of the spectrum as
\begin{equation} \label{eq:eigenspectrum}
    \widehat{S}_{xx}^{(k)}(f) = |\widetilde{x}^{(k)}(f) |^2.
\end{equation}



The multitapers do more than just make it possible to estimate the MSC: the averaging reduces the variance of the overall estimator and the use of $K$ orthogonal tapers reduces the bias. \cite{bt} provide a comprehensive overview of tapering and quantify the bias reduction of different types of tapers, while \citet{harris78} reviews common choices of taper including the popular Hamming and Hann windows.
The multitapers are the discrete prolate spheroidal sequences (DPSS) discovered by \citet{s78} and extended to time series with unequal sampling cadence by \citet{bronez1988spectral}. In a recent paper by \citet{chave2019multitaper}, the missing-data case was shown to have a straightforward set of tapering functions (``missing-data Slepian sequences'' or MDSS). The MDSS and DPSS are defined by the eigenvalue equation \citep{H2021}
\begin{equation} \label{eq:mdss} 
  \lambda^{(k)} \sum_{m=0}^{N-1} \frac{w_n^{(k)}}{(t_n-t_m)} =
  \sum_{m=0}^{N-1} \frac{\sin (2\pi \varpi [t_n - t_m]) }{\pi [t_n - t_m]} w_m^{(k)},
\end{equation} 
where the $t_n$ are the timestamps, which may have missing values (e.g. $t_n = 0, 1, 2, 5, 6, 8, \ldots, M$). In Equation \eqref{eq:mdss}, $\varpi$ is the user-defined {\it bandwidth} of the multitaper spectral window and each eigenvalue in the range $0 < \lambda^{(k)} < 1$ is the {\it spectral concentration} of the associated taper $w_n^{(k)}$. The spectral concentration measures the fraction of power at frequency $f_0$ that the multitaper power spectrum estimator keeps contained inside the band $-\varpi \leq f_0 \leq \varpi$. 

To estimate the bias of a power spectrum estimator, we follow Appendix D of \cite{scargle1982studies}. We define each taper's \emph{pseudowindow} at frequency $f_0$, $W^{(k)}_{f_0}(f)$, by substituting the deterministic sequence $a_n(f_0)$, which is a single sinusoid with frequency $f_0$, in for $x_n$ in Equation \eqref{eq:taperedNUFFT}:
\begin{align}
    a_n(f_0) &= \cos (2 \pi f_0 n); \\
    W^{(k)}_{f_0}(f) &= \left| \widetilde{a}^{(k)}(f) \right|^2.
    \label{eq:pseudowindow}
\end{align} 
When there are no missing data, 
the bias is given by the standard convolution formula:
\begin{align}\label{eq:nomissingbias}
\operatorname{bias} \{\widehat{S}^{(k)}_{xx}(f)\} &= \mbox{E} \{ \widehat{S}^{(k)}_{xx}(f) - S(f) \} \\
\mbox{E} \{ \widehat{S}^{(k)}_{xx}(f) \} &= (W^{(k)}_0 * S) (f).
\end{align}
That is, the expected value of the power spectrum estimate is the pseudowindow at zero frequency $W^{(k)}_0(f)$ (hereafter simply referred to as the spectral window) convolved with the true power spectrum that we are trying to estimate, $S(f)$. If there existed a finite data taper for which $W^{(k)}_0 (f) = \delta(f)$, then we could construct an estimator for the spectrum with zero bias. However, such a taper does not exist due to uncertainty principles. \cite{s78} was able to give a set of data tapers that have Fourier transforms which concentrate most of their mass under $(-\varpi, \varpi)$, thus approximating a delta function. When the time sampling is even, one can use the Cauchy inequality to put an upper bound on the broadband bias of the multitaper power spectrum estimators. Using $\widehat{s}_x^2 = (N-1)^{-1}\sum_n (x_n - \bar{x})^2$ to denote the sample variance of the time series, 
\begin{equation}
    \mbox{E} \left \{ \widehat{S}^{(k)}_{xx}(f) - S(f) \right \} \leq (1 - \lambda^{(K-1)}) \widehat{s}^2_x.
    \label{eq:biasbound}
\end{equation}

When the time indices
contain missing values, the use of the nonuniform Fourier transform means the bias no longer has the convenient form of a convolution, as in \eqref{eq:nomissingbias}. Instead, the pseudowindow is different at every choice for the input sinusoid frequency $f_0$. The pseudowindow is useful for one main reason: by definition, it is the response of one's chosen power spectrum estimator to an input cosine sequence on the missing-data timestamps. This is to be interpreted the same way as the spectral window $W^{(k)}_0(f)$: we would like for each pseudowindow to be as delta-like as possible so as to faithfully reflect the shape of the true power spectrum. When the gaps introduced by missing data are not overly small and numerous, 
we compute $W^{(k)}_0(f)$ and use its shape as a non-quantitative heuristic for the signature of a sinusoid at any frequency. However, when the missing data indices are such that they could bias the spectrum at high frequency, it is best to compute the frequency-dependent pseudowindows so as to avoid making the wrong inference. Unfortunately in our experience, and in the example we show in this paper, Lomb-Scargle pseudowindows can be biased by \emph{sidelobes}, or peaks adjacent to the center frequency, that have large magnitude. Some Lomb-Scargle pseudowindows even have large far-field satellite peaks that appear due to repeating patterns in the missing data indices. These sidelobes and satellite peaks can \emph{grossly} misrepresent the shape of the true power spectrum and potentially lead to false inferences.


An important consideration for the estimation of the multitaper spectrum is the selection of the bandwidth parameter $\varpi$ \citep{haley17}. We give a discussion in \citet[\S 2.7]{missingdata} of how to select the appropriate bandwidth for the science goal, but it is generally advisable to set the bandwidth so that it is an integer or half-integer multiple of the fundamental resolution unit, the Rayleigh resolution, equal to the reciprocal of the duration of the time series (recall that our time series have duration $M$). This ensures that a strong sinusoidal component in the time series will produce a high power spectrum peak twice as wide as the bandwidth, or $2M\varpi$ Rayleigh resolutions wide. For time series with $N \sim 1000$ and $N \lessapprox M$, $M\varpi = 4$ is often a good starting point.

While the multitaper method's bias-reduction ability comes from the shapes of the DPSS spectral windows, its variance suppression is the result of averaging together the $K$ independent estimators $\widehat{S}^{(k)}_{xy}(f)$. Multitapering reduces the variance in the resulting power- and cross-spectrum estimates by a factor of nearly $K$. Equation \eqref{eq:mdss} yields at most $K = \lfloor 2M\varpi \rfloor$ useful, highly concentrated tapers with $\lambda \simeq 1$ (where $\lfloor \cdot \rfloor$ represents the floor function); for $(k) > 2 M \varpi$ the spectral concentration drops precipitously and the shape of $W^{(k)}_0(f)$ no longer approximates a delta function. There is therefore a tradeoff between bias and variance: higher values of $\varpi$ provide more tapers and therefore improve variance suppression, but also admit more local bias by widening the spectral window (see \citet{thomson2014spacing}, \citet{SDR2022}, \S 3, and \citet{missingdata} for more about bias). Reducing $\varpi$ narrows the spectral window, improving local bias, but also reduces the maximum possible value of $K$.

\section{The jackknifed multitaper MSC estimator}
\label{sec:jackcoh}

The most basic multitaper power spectrum estimator is a sample average of the $K$ sequences of squared eigencoefficients: $\widehat{S}_{xx}(f) = (1/K) \sum_{k=0}^{K-1} |\widetilde{x}^{(k)}(f)|^2$. However, coherence estimation is done using the {\it jackknife}
mean, which removes a first-order bias term in the MSC variance estimator \citep{thomson91}. The following use of jackknifing is based on the simplest ``leave-one-out'' paradigm \citep{efron82}. For example, one could calculate $N$ leave-one-out averages of the flux $F$ of a star observed $N$ times using $\widehat{F}^{\setminus n} = (1 / [N-1]) \sum_{m = 0, m \neq n}^{N-1} F_m$. Here we will jackknife over the multitapers to compute $K$ leave-one-out estimates of each quantity in Equation \ref{eq:cohest}. We will apply the \citet{fisher29} scaled inverse hyperbolic arctangent transformation,
\begin{equation}
    z(f) = \sqrt{2K-2} \operatorname{atanh} \left (\widehat{C}^2_{xy}(f) \right),
    \label{eq:Fishertrans}
\end{equation}
while jackknifing in order to stabilize the variance of the MSC estimator. In order to compute the jackknife variance estimate, it is assumed that the individual eigenestimates are normally distributed. The Fisher transformation brings the distributions of statistics bounded between zero and one closer to Gaussian \citep[\S 5.3]{Miller74}. The estimated variance of $z(f)$ then follows a Student-$t$ distribution which can be used to construct a confidence interval for the transformed MSC estimate. 

The leave-one-out estimators of the transformed MSC are
\begin{equation} \label{eq:jkMSCQ}
z^{\setminus m}(f) = \sqrt{2K-2} \operatorname{atanh} \left ( \frac{ \left| \sum_{\substack{k = 0 \\ k \neq m}}^{K-1} 
\widetilde{x}^{(k)} (f) \widetilde{y}^{*(k)}(f) \right|}{\sqrt{ \sum_{\substack{k = 0 \\ k \neq m}}^{K-1} | \widetilde{x}^{(k)}(f)|^2 \cdot 
\sum_{\substack{k = 0 \\ k \neq m}}^{K-1} | \widetilde{y}^{(k)}(f)|^2 }} \right),
\end{equation}
where the quantity in large brackets estimates the absolute coherence before squaring. Let $z^{\mbox{all}}(f)$ denote Equation \eqref{eq:jkMSCQ} evaluated using all of the tapers (no deletion). Each $z^{\setminus m}(f)$ is associated with a {\it pseudovalue} $z_m(f)$ given by 
\begin{equation} \label{eq:JkmeanCoh}
  z_m(f) = K z^{\mbox{all}}(f) - (K-1) z^{\setminus m}(f).
\end{equation}
The average of the pseudovalues
\begin{equation}
    z(f) = \frac{1}{K} \sum_{k=0}^{K-1} z_m(f)
    \label{eq:zest}
\end{equation}
is the estimator for transformed MSC. 

To construct confidence intervals based on the Student-$t$ distribution, we require an estimate of the variance of $z(f)$. We take that estimate from the variance of the pseudovalues,
\begin{equation}
  \widehat{\operatorname{Var}} \{ z^{\mbox{all}}(f) \} = \frac{1}{K(K-1)} \sum_{m=0}^{K-1} \left[z_m(f) - z(f) \right]^2,
\end{equation}
since $\mbox{E} \{ \operatorname{Var} \{ z^{\mbox{all}}(f) \} \} = \mbox{E} \{ \operatorname{Var} \{ z(f) \} \}$. The confidence interval for the transformed MSC is then
\begin{equation} \label{eq:CohJKCI}
\frac{z(f) - t_{K-1}(1-\alpha/2)\sqrt{\widehat{\operatorname{Var}} \{ z^{\mbox{all}}(f) \}}}{\sqrt{K-2}} < z <
\frac{z(f) + t_{K-1}(1-\alpha/2)\sqrt{\widehat{\operatorname{Var}} \{ z^{\mbox{all}}(f) \}}}{\sqrt{K-2}},
\end{equation}
where $t_{K-1}(1-\alpha/2)$ denotes the Student-$t$ critical value for the risk level $0<\alpha<1$ (significance level $1-\alpha$) and $K-1$ degrees of freedom. Confidence intervals for the MSC mean the following: given an infinite number of pairs of realizations of the time series $x_n$ and $y_n$, the computed value for the transformed MSC ought to fall inside the confidence interval $100(1-\alpha)$\% of the time.

In the next section, we use two simultaneous, missing data time series to demonstrate the mathematical operations in \S \ref{sec:explaincoherence}---\ref{sec:jackcoh}.

\section{Missing-data multitaper MSC between solar Lyman $\alpha$ and disturbance storm time index}
\label{sec:samplecalc}


Here we illustrate how to estimate multitaper magnitude-squared coherence using daily average measurements of the solar Lyman $\alpha$ index \citep[Ly$\alpha$,][]{machol19} and the Disturbance Storm Time index (Dst) defined by \citet{sugiura64}, which measures perturbations to the horizontal component of the geomagnetic field. We first describe the characteristics of each time series (\S \ref{subsec:sampledata}), then present the set of tapers we use for the MSC estimate (\S \ref{subsec:sampletapers}) and the power spectra of the two time series (\S \ref{subsec:spectra}). We discuss the MSC estimate between Ly$\alpha$ and Dst in \S \ref{subsec:msc_lya_dst}, focusing on known mid-term periodicities that are jointly traced by both time series.

\subsection{A multitaper demonstration dataset}
\label{subsec:sampledata}

\begin{figure}
    \centering
    \includegraphics[width=\textwidth]{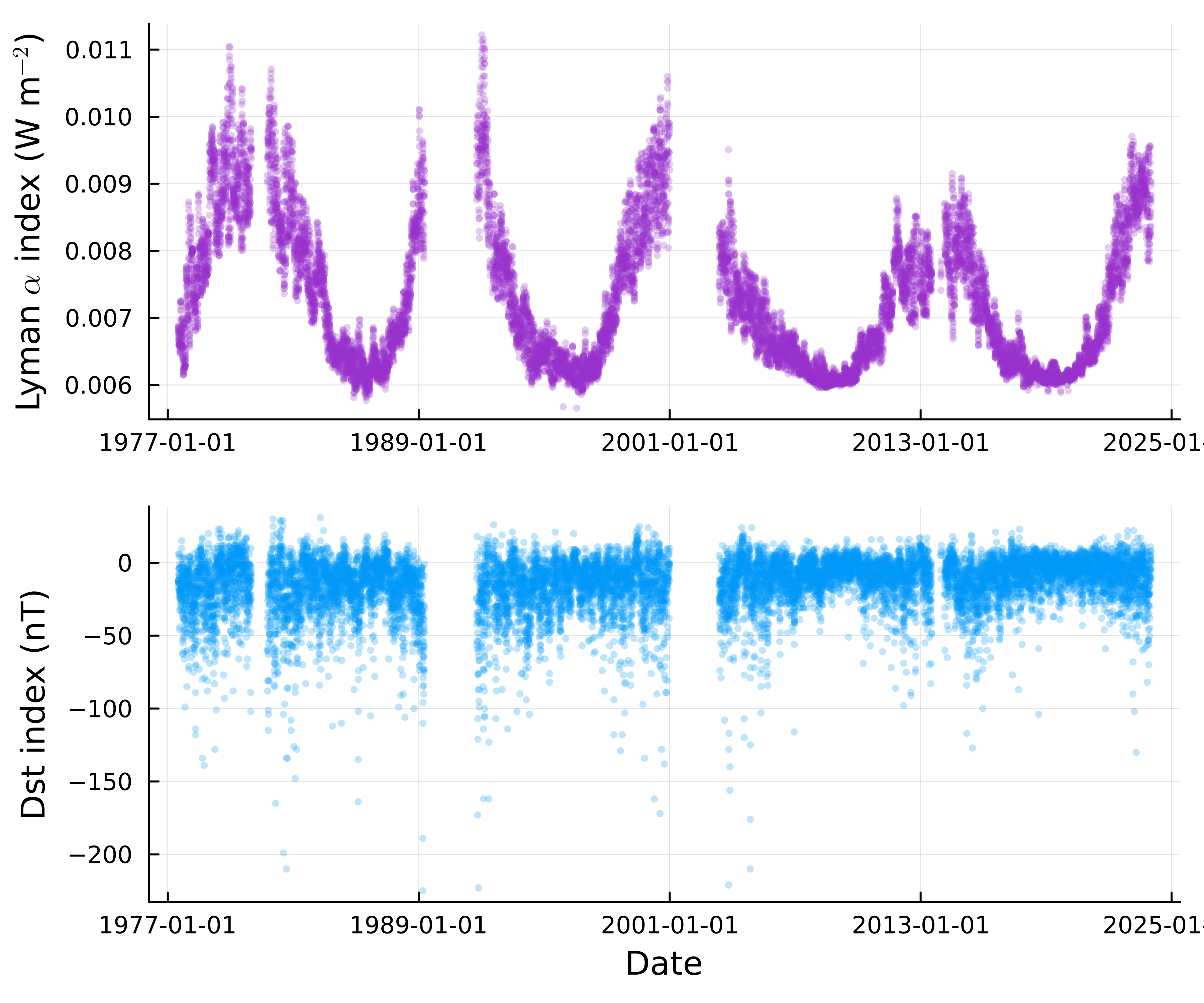}
    \caption{{\bf Top:} Daily average Ly$\alpha$ index measured from satellites in low-Earth orbit \citep{machol19}. {\bf Bottom:} Daily average Dst index measured concurrently with the Ly$\alpha$ satellite data.}
    \label{fig:Lya_Dst}
\end{figure}

The Ly$\alpha$ index is the 121--122~nm daily average solar irradiance incident at 1~au, measured since 2 July 1977 by a series of satellites in Earth's upper atmosphere \citep{woods00, machol19}. The reference irradiance scale is established by the SOLSTICE instrument aboard the SORCE satellite, which was launched into low-Earth orbit in 2003 \citep{mcclintock05a, mcclintock05b}. Data from the Atmospheric Explorer E \citep{hinteregger81}, Solar Mesosphere Explorer \citep{barth83}, Upper Atmosphere Research Satellite \citep{rottman93, woods96}, and Geostationary Operational Environmental Satellites 17 and 18 \citep{thiemann19} are scaled to match the SOLSTICE SORCE data \citep[\S 3]{machol19}. The latest version of the Ly$\alpha$ composite uses a model based on the Bremen Mg~II index \citep[$\lambda = 280$~nm,][]{snow14} to fill in the time series when Ly$\alpha$ observations are unavailable due to gaps between satellite missions. The Ly$\alpha$ time series is also extrapolated back to 1947 based on solar radio fluxes at $\lambda = 10.7$~cm \citep{tapping13} and $\lambda = 30$~cm \citep{dudokdewit17}. The resulting composite is an excellent resource for investigations of the Schwabe/Hale cycle variability and its interaction with other periodicities, such as the quasibiennial oscillation \citep[e.g.][]{mursula03}. However, no transformation from one observable to another is perfect, so the model-based Ly$\alpha$ data may be vulnerable to systematics. Some observers might therefore prefer to use only the irradiances calculated directly from satellite data. Doing so creates a missing data time series with underlying uniform observing cadence (Figure \ref{fig:Lya_Dst}, top panel) whose MSC with a simultaneous time series can be calculated following \citet{H2021}.

The Dst index records the departure of the horizontal component of the geomagnetic field $H$  from the previous four years' baseline level $H_{\rm base}$. $H$ is sensitive to the equatorial east-west ring current in the inner magnetosphere, which is roughly axisymmetric during magnetically quiet times \citep{ganushkina17}. During a geomagnetic storm, plasma transport generates asymmetries in the ring current, creating a southward perturbation field $\Delta H$ \citep[e.g.][]{mitchell01}. According to \citet{sugiura64} and \citet{dst}, Dst calculation proceeds as follows: At each of four low-latitude geomagnetic observatories, the locally measured horizontal field components $H_o$ during the five most magnetically quiescent days from each month are averaged to form an annual mean for the current year plus each of the previous four calendar years. Each observatory's time-varying model for the baseline field $H_{\rm base}$ is then
\begin{equation}
    H_{\rm base,o}(\tau) = A + B \tau + C \tau^2,
    \label{eq:Hbase}
\end{equation}
where $\tau$ is the time elapsed from a reference epoch. The next step is to model out the annual and diurnal variations in $H$ caused by the quiet sun, $S_q$. The annual model is calibrated to the average values of $H$ from each month's five quietest days based on aggregated data from geomagnetic observatories worldwide. Each observatory's diurnal variation in $S_q$ in a given month is calculated based on hourly average $H_o$ measurements from the month's five locally quietest days that most overlap with the five worldwide quiet days selected for the annual model. The observatory-specific solar-quiet model $S_{q,o}$(t) is then determined by a double Fourier series fit to the quiet-sun averages:
\begin{equation}
    S_{q,o} = \sum_{i=1}^6 \sum_{j=1}^6 A_{ij} \cos(ih + \alpha_i) \cos(js + \beta_j),
    \label{eq:solarquiet}
\end{equation}
where $h$ is the hour in UT and $s$ is the month number. Finally, the global average Dst index is
\begin{equation}
    \operatorname{Dst} = \frac{\sum_{o=1}^4 {H_o - H_{\rm base,o} - S_{q,o}}}{\sum_{o=1}^4 \cos \varphi_o},
\end{equation}
where $\varphi_o$ is the dipole latitude of the observatory. 
Quiet periods in which the ring current is minimally disturbed have $\operatorname{Dst} > -30$~nT \citep{kim05}. During the 1989 magnetic superstorm that shut down the Quebec power grid, the Dst index plummeted to $-589$~nT \citep{boteler19}. According to \citet{cannon13}, a magnetic superstorm would disable global positioning satellites for 1--3 days and expose airplane passengers and crew to enough high-energy radiation to increase their lifetime risk of fatal cancer by 25\%.

We create simultaneous Dst and Ly$\alpha$ time series by selecting the daily average Dst values from dates that also have Ly$\alpha$ spacecraft observations. The bottom panel of Figure \ref{fig:Lya_Dst} shows the Dst time series used in this demonstration.

\subsection{A sample set of tapers}
\label{subsec:sampletapers}

\begin{figure}
    \centering
    \includegraphics[width=\textwidth]{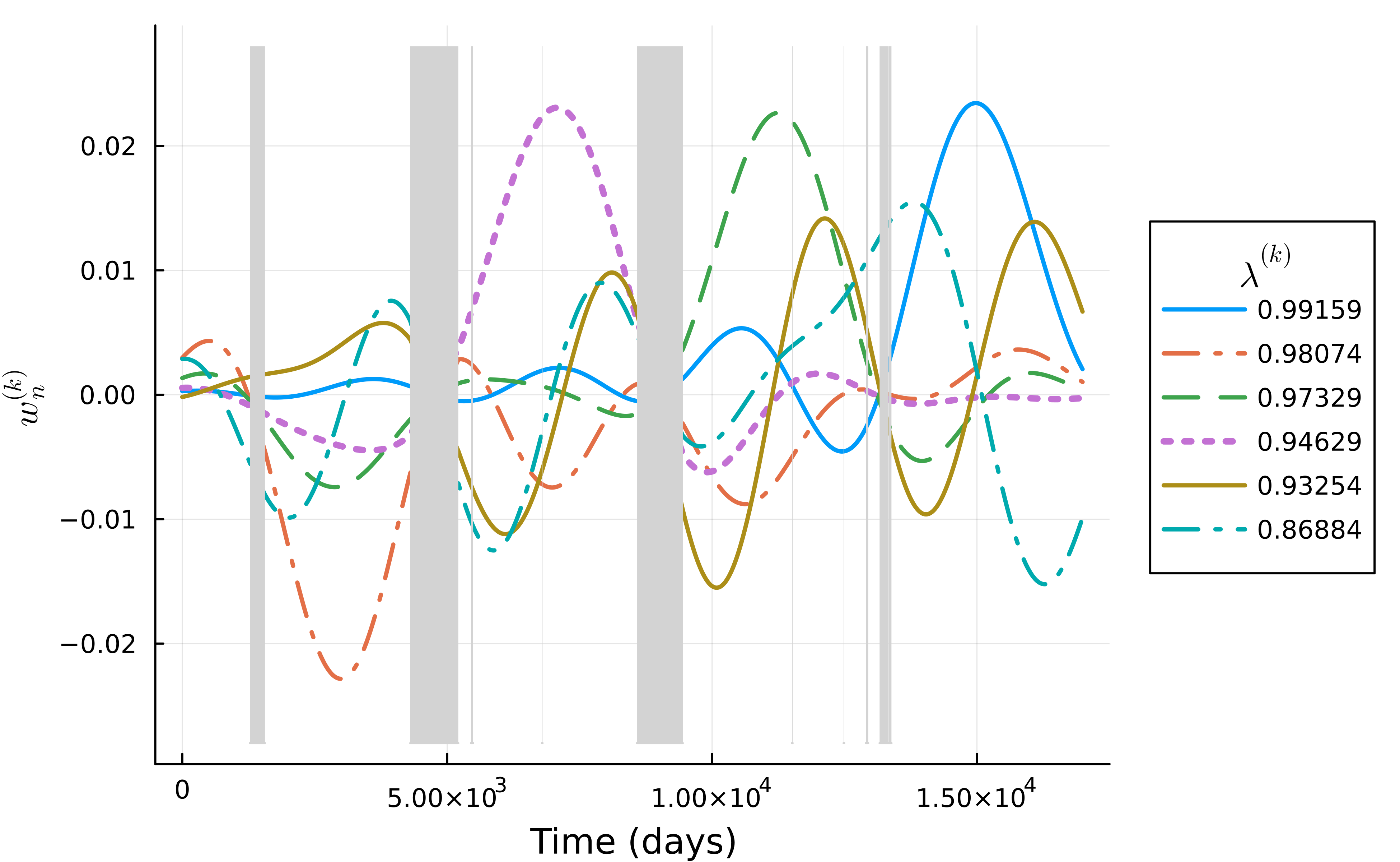}
    \caption{Multitapers calculated with $N\varpi = 4.5, K = 6$ for the timestamps of the simultaneous solar Ly$\alpha$ and Dst index measurements shown in Figure \ref{fig:Lya_Dst}. The legend shows the spectral concentration $\lambda^{(k)}$ of each taper. Gray shading indicates gaps in the time series.}
    \label{fig:Lya_Dst_tapers}
\end{figure}

For the Ly$\alpha$-Dst multitaper power spectrum and MSC estimates, we select an approximate time-bandwidth product $M\varpi = 4.5$ and use $K = 6$ tapers. The bandwidth is $\varpi = 0.00031$~day$^{-1}$. Figure \ref{fig:Lya_Dst_tapers} shows the tapers calculated using Equation \eqref{eq:mdss}. Our bandwidth choice is set primarily by resolution constraints: an observer studying the sun's midterm periodicities would want to be able to distinguish between (for example) the 1.3-year rotation period at the base of the convection zone measured from GONG data \citep{howe00} and the 1.6-year cycle observed in the Bremen composite Mg~II index \citep{mehta22}. However, \citet{haley17} present an algorithm for finding the multitaper bandwidth that minimizes the mean-squared error of the power spectrum estimate.
With $M\varpi = 4.5$, we could use up to $K = 9$ tapers, but we choose to leave out $w^{(6)}_t$, $w^{(7)}_t$, and $w^{(8)}_t$ because they have lower spectral concentrations: $\lambda = 0.84, 0.80, 0.56$ respectively. 

In general, if $N$ and $\varpi$ are held constant, spectral concentrations decrease rapidly as one adds ever larger numbers of brief gaps in the time series. Similarly, large data gaps tend to reduce the information we have about quickly varying periodicities. In addition, there might even be artificial periodicities generated by a quasi-periodic occurrence of data-gaps (for example, Fig. \ref{fig:Lya_Dst} shows the presence of missing data during each solar maximum: 1979, 1989, 2001, 2014). These effects can be investigated (but not eliminated) by close examination of the pseudowindows at the frequencies of interest. Finally, when the MSC is being estimated, there is also a well known biasing due to \emph{misalignment}: large time delays between series can cause rapidly varying phase \cite[Section III-D]{carter1987coherence} and can result in underestimating the MSC. When data gaps occur, misalignment can be exacerbated to the point of missing even strongly coherent signals.

\subsection{Missing-data multitaper spectra of solar Ly$\alpha$ and Dst}
\label{subsec:spectra}

\begin{figure}
    \centering
    \includegraphics[width=\textwidth]{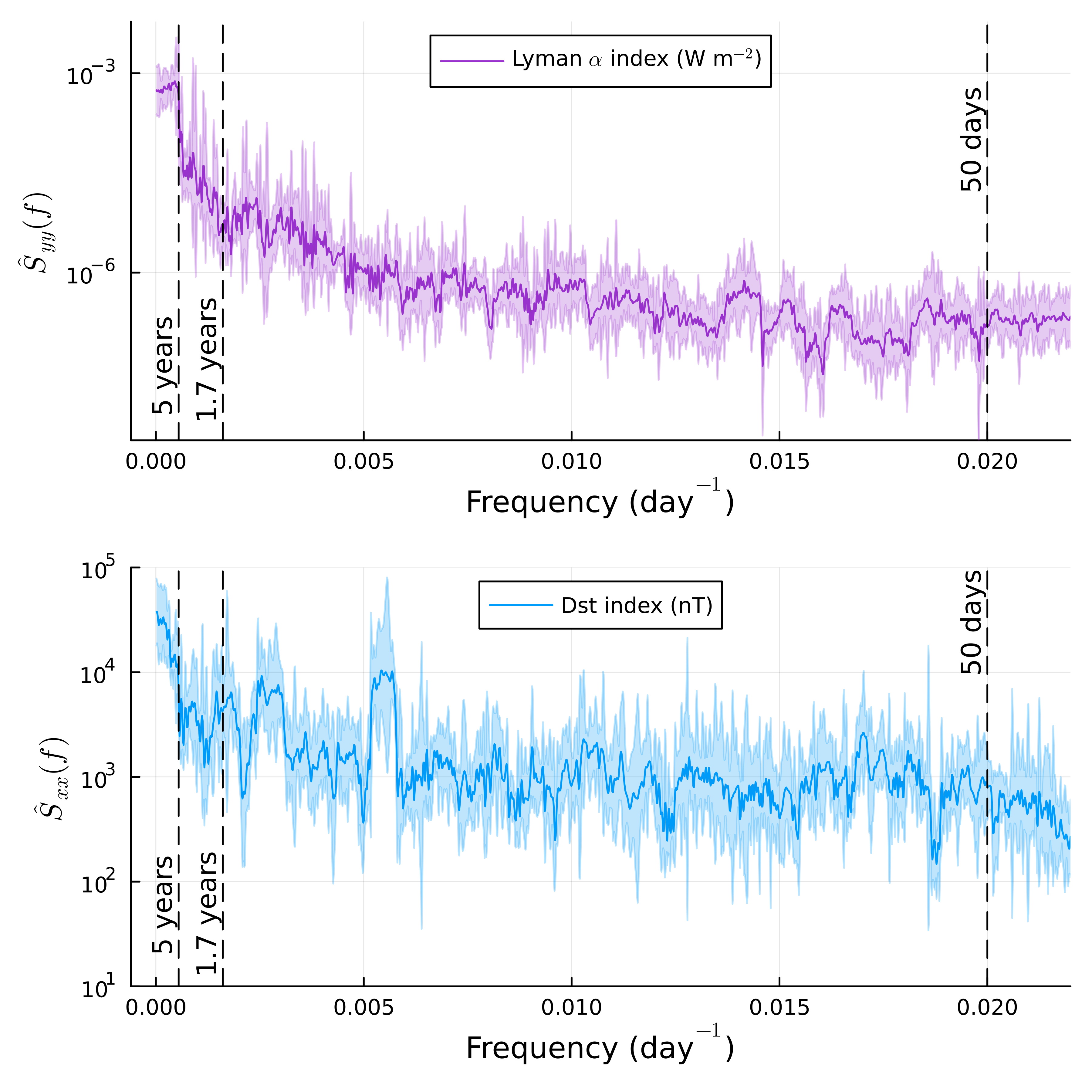}
    \caption{{\bf Purple curve:} Missing data multitaper spectrum of daily average Ly$\alpha$ index. {\bf Blue curve:} Spectrum of daily average Dst index measured concurrently with the Ly$\alpha$ satellite data. Shading represents 95\% confidence intervals for the log-spectrum using the jackknife technique. Vertical lines indicate frequencies that correspond to periods of 5 years, 1.7 years, and 50 days. Original time series were shown in Fig. \ref{fig:Lya_Dst}.}
    \label{fig:Lya_Dst_MDSpec}
\end{figure}

Fig.\ \ref{fig:Lya_Dst_MDSpec} shows the power spectra of the Dst and Ly$\alpha$ series, computed using the missing data Slepian sequences shown in Fig. \ref{fig:Lya_Dst_tapers} and the missing data multitaper method described in \cite{chave2019multitaper}. That is, 
\begin{equation}
    \widehat{S}_{xx}(f) = \frac{1}{K} \sum_{k=0}^{K-1} \widehat{S}_{xx}^{(k)}(f) 
\end{equation}
where $\widehat{S}^{(k)}(f)$ is given in \eqref{eq:eigenspectrum}. Note that the Dst spectrum is everwhere several decades larger than Ly$\alpha$, but Ly$\alpha$ has somewhat greater dynamic range and therefore is more susceptible to bias from spectral leakage. The multitapers shown in Fig.\ \ref{fig:Lya_Dst_tapers}, however, have concentrations ranging from $0.869 - 0.991$, indicating that bias introduced in the high-frequency range by large power at low frequencies will be minimal.

In the next section, we will describe how to use the multitapers to estimate MSC. 

\subsection{Magnitude-squared coherence between solar Ly$\alpha$ and Dst}
\label{subsec:msc_lya_dst}

\begin{figure}
    \centering
    \includegraphics[width=\textwidth]{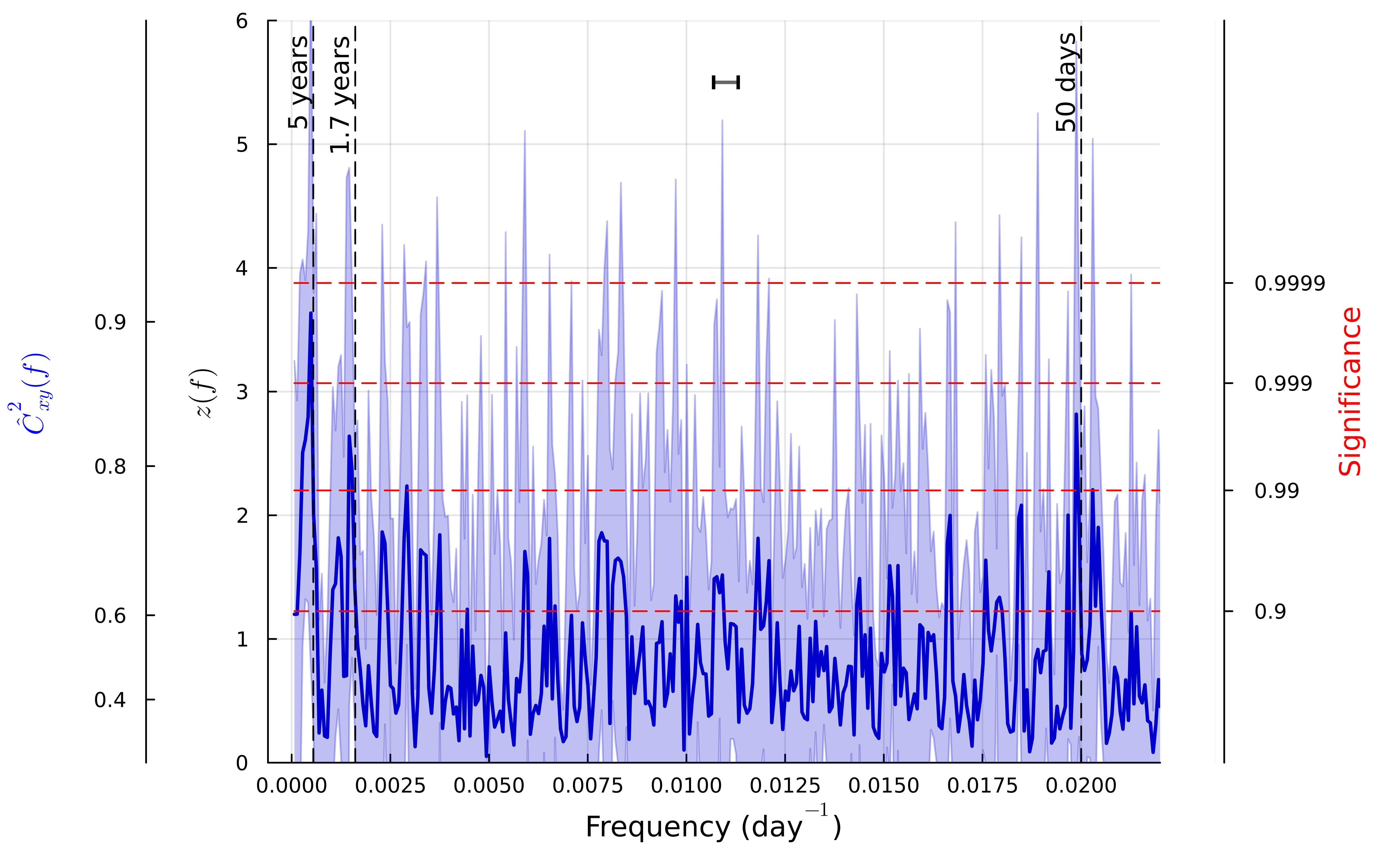}
    \caption{Magnitude-squared coherence at mid-term frequencies between the gapped Ly$\alpha$ and Dst time series pictured in Figure \ref{fig:Lya_Dst}. The leftmost y-axis records the MSC estimate $\widehat{C}_{xy}(f)$, while the middle y-axis shows the Fisher transformed version $z(f)$. The right-hand y-axis indicates the statistical significance level, with horizontal red lines marking 90.00\%, 99.00\%, 99.90\%, and 99.99\%. Light blue shading shows the 95\% confidence interval from Equation \eqref{eq:CohJKCI} and the horizontal black bar shows full bandwidth $2 \varpi$, which is the resolution unit of $z(f)$.}
    \label{fig:Lya_Dst_coh}
\end{figure}

Since particles emitted by the sun interact with the geomagnetic field, we expect to detect oscillations that are jointly traced by Ly$\alpha$ and Dst. Here we focus on mid-term periodicities that fall between the Schwabe cycle (11 years) and the synodic rotation period (27.3 days). In Figure \ref{fig:Lya_Dst_coh}, we show the MSC between the simultaneous solar Ly$\alpha$ and Dst time series plotted in Figure \ref{fig:Lya_Dst} for $f \leq 0.02$~day$^{-1}$ ($P > 50$~days). From left to right, the three vertical axes indicate the MSC estimate (Equation \ref{eq:JkmeanCoh}), the Fisher-transformed estimate $z(f)$ (Equation \ref{eq:zest}), and the statistical significance level (equivalent to $1 - \operatorname{FAP}$, where FAP is the false alarm probability). Three shared oscillatory signals that have $> 99$\% statistical significance and frequencies within one half-bandwidth $\varpi$ of known physical processes are marked with dashed vertical lines. Figure \ref{fig:Lya_Dst_MDSpec} shows that each coherent oscillation frequency corresponds to a local maximum in either the Ly$\alpha$ or the Dst power spectrum.

One well-studied oscillation that manifests in Figure \ref{fig:Lya_Dst_coh} has period 1.7~years (621 days, $f = 0.00161$~day$^{-1}$) and belongs to a class of quasibiennial oscillations (QBOs) that have periods 1--4 years \citep[see review by][]{norton23}. The 1.7-year periodicity occurs intermittently, and is part of a process where the dominant approximate period alternates between 1.2--1.4 years and 1.5--1.7 years \citep{mursula03}. A 1.7-year periodicity has been detected in numerous solar and heliospheric observables, including coronal hole area, cosmic ray intensity \citep{valdesgalicia96, mavromichalaki03, lopezcomazzi23}, interplanetary proton flux \citep{laurenza09}, northern hemisphere sunspot group number \citep{mendoza11}, composite solar flare index \citep{velascoherrera22}, and Bremen composite Mg~II index \citep{mehta22}.
According to \citet{hiremath10}, the the 1.5--1.7-year quasiperiodicity may result from Alfven wave perturbations to the toroidal magnetic field structure.
The 1.7-year cycle can also be seen in geomagnetic diagnostics such as Ap index \citep{tsichla19}
and the extended aa index defined by \citet{mursula01}. 
Since the Ly$\alpha$ and Dst data underlying Figure \ref{fig:Lya_Dst_coh} probe solar and geomagnetic behavior, respectively, it is no surprise that they jointly record a 1.7-year quasiperiodicity.

Figure \ref{fig:Lya_Dst_coh} also records a 5-year periodicity that matches a known medium-term quasiperiodic oscillation with a period of 4--5.5 years (1460--2010 days, 0.000498--0.000684~day$^{-1}$). This cycle has been identified in solar metrics such as sunspot number \citep{singh14, ravindra22}, North-South asymmetry index \citep{depaula22}, sunspot umbra/penumbra ratio \citep{chowdhury24}, polar faculae counts \citep{deng14}, solar wind speed \citep{prabhakarannayar02}, and plage area, F10.7, and the coronal index \citep{elborie20}. It also appears in the geomagnetic aa index \citep{singh14} and Ap index \citep{frasersmith72, prabhakarannayar02}. As with the 1.5--1.7-year cycle, \citet{hiremath10} associates the 4--5.5-year cycle with perturbations to the toroidal magnetic field, while \citet{sokoloff20} favor nonlinear saturation of the solar dynamo. Figure \ref{fig:Lya_Dst_coh} shows that the statistical significance of the MSC between Ly$\alpha$ and Dst in the 0.000498--0.000684~day$^{-1}$ band exceeds the 99.9\% level. At $f = 0.000517$~day$^{-1}$, the bottom of the 95\% confidence interval exceeds the 90\% significance level, indicating an extremely low probability that the oscillation detection is a false positive.

The final mid-term periodicity that reaches the 99\% significance level in Figure \ref{fig:Lya_Dst_coh} has a period of 50~days, which is consistent with previous detections of quasiperiods between 50 and 57 days ($f = 0.0175$--0.0200~day$^{-1}$). \citet{bai03} identified a statistically significant 51-day signal in a normalized Rayleigh spectrum of the daily solar flare occurrence rate from Cycle 19, while \citet{deng14} found the same quasiperiod in polar faculae counts. The geomagnetic Ap index also records a 51-day period \citep{singh17}. A multitaper analysis by \citet{kilcik10} that used sinusoidal tapers instead of the DPSS/MDSS described in \S \ref{sec:multitaper} (Equation \ref{eq:mdss}) found a similar periodicity, 53 days, in the solar flare index. \citet{chowdhury09} found a significant oscillations at 54 and 57 days in whole-disk sunspot area measurements from Cycle 22. Notably, 54 days is a subharmonic of the sun's synodic rotation period, while 51 days is a subharmonic of the sidereal rotation period. In a multitaper power spectrum of the daily F10.7 index values from 1965--2014 computed with the standard DPSS tapers, \citet{roy19} discovered a significant 57-day periodicity. It is unclear whether the ensemble of 50--57-day signals is associated with rotation, or whether all of the reported signals belong to the same physical process.


One mid-term oscillation that we do not detect in the Ly$\alpha$--Dst MSC is the 154-day signal ($f = 0.00650$~day$^{-1}$) first found in the occurrence rate of hard solar flares by \citet{rieger84}. A family of similar periods has since been found in numerous solar and geomagnetic indices \citep[e.g.][]{norton23}. Proposed explanations for the Rieger periodicities include harmonic(s) of the QBO \citep{krivova02}, outbreaks of active regions on the solar surface \citep{vecchio12}, and tachocline nonlinear oscillations \citep{dikpati05, dikpati21}. 
Based on a wavelet analysis of sunspot data from cycles 14--24, \citet{gurgenashvili16} suggest that Rieger-type periodicities are present in all phases of all solar cycles, but that period is a function of cycle strength: the characteristic weak-cycle Rieger period of 185--195 days drops to 155--165 days during strong cycles. Since multitaper MSC is a Fourier analysis technique, it is best suited for identifying shared oscillations without substantial period drift. A wavelet coherence analysis similar to that of \citet{volvach24} would be a better test of whether Ly$\alpha$ and Dst jointly trace the Rieger family of periodicities. 

\section{Conclusions}
\label{sec:conclusions}

In this paper, we have demonstrated how to use multitaper magnitude-squared coherence to diagnose oscillations in common to two time series that have underlying uniform observing cadence but are missing some data. With multitaper, the observer can shape the spectral window to suppress bias, average together multiple independent Fourier spectrum estimates to reduce variance, and compute confidence intervals on the MSC estimate. For gapped time series, the \citet{chave2019multitaper} and \citet{H2021} multitaper method is superior to the Welch’s method-based MSC estimator demonstrated by \citet{SDR2022} because it optimally balances bias, variance, and bandwidth: holding two of the three metrics constant, multitaper produces a lower value of the third than any other frequency-domain technique \citep{bronez92}. However, Welch’s method is more general and can be applied to time series with uneven observing cadence.

Applying the missing-data multitaper MSC estimator to
simultaneous daily-average Ly$\alpha$ and Dst time series that
have missing data due to gaps between Ly$\alpha$-measuring
satellite missions, we detect three well-known midterm solar
oscillations with statistical significance $>99$\%. Two of the
shared oscillations, those with periods of 5~years and 1.7~years, may probe perturbations to the toroidal component of the sun's magnetic field \citep{howe00, hiremath10}. The 50-day signal may be associated with a rotation subharmonic. The knowledge that two very different observables---an ultraviolet absorption line flux and a measure of Earth's east-west magnetospheric ring current---jointly trace multiple physical processes is a useful constraint on models of the solar dynamo and solar wind transport. 

For future work, phase spectra (Equation \ref{eq:phasespec}) from time series with shorter observing cadence could be used to search for lead-lag relationships. If Ly$\alpha$ does not affect the Earth's magnetosphere directly, but simply co-varies with solar wind particle fluxes, then we expect Dst to lag Ly$\alpha$ by approximately the amount of time it takes the solar wind to travel to Earth. On the other hand, if Ly$\alpha$ photons themselves influence the magnetosphere---for example, by increasing the atmospheric ionization rate---then the lag between Ly$\alpha$ and Dst should be of order the sun--Earth light travel time. The multitaper cross-spectral analysis techniques developed by \citet{H2021} and demonstrated here can be applied to a wide range of solar and heliospheric physics problems.

\section*{Open Research Section}

\section*{Data availability}

The Ly$\alpha$ time series used in \S \ref{sec:samplecalc} 
was obtained from the Lyman $\alpha$ composite served by the Laboratory for Atmospheric and Space Physics Interactive Solar Irradiance Datacenter (LISIRD, \url{https://lasp.colorado.edu/lisird/data/composite_lyman_alpha}). 
Daily averaged Dst data were downloaded from NASA OMNIWeb (\url{https://omniweb.gsfc.nasa.gov}). Both downloads are archived at \citet{zenodo}. 

\section*{Software availability}

The \texttt{Jupyter}/\texttt{Julia 1.11.4} notebook used to analyze the data in \S \ref{sec:samplecalc} is archived at \citet{zenodo}. 
All multitaper calculations use the \texttt{Multitaper.jl} software package, which is described in \citet{multitaperpkg} and hosted at \url{https://github.com/lootie/Multitaper.jl}. The \texttt{Multitaper.jl} GitHub repository has has tutorial notebooks on multitaper power spectrum and MSC estimation for time series with and without gaps. For users who prefer \texttt{python} to \texttt{Julia}, \texttt{Multitaper.jl} has a python wrapper, \texttt{multitaperpy} (https://github.com/lootie/multitaperpy/).

\acknowledgments

Funding for this work was provided by National Science Foundation grant 2307978 and by the U.S.\ Department of Energy, Office of Science, Advanced Scientific Computing Research, under contract number DE-AC02-06CH11357.

The submitted manuscript has been created by UChicago Argonne, LLC, Operator of Argonne National Laboratory (“Argonne”). Argonne, a U.S. Department of Energy Office of Science laboratory, is operated under Contract No. DE-AC02-06CH11357. The U.S. Government retains for itself, and others acting on its behalf, a paid-up nonexclusive, irrevocable worldwide license in said article to reproduce, prepare derivative  works, distribute copies to the public, and perform publicly and display publicly, by or on behalf of the Government. The Department of Energy will provide public access to these results of federally sponsored  research in accordance with the DOE Public Access Plan. http://energy.gov/downloads/doe-public-access-plan

%
%

\bibliography{nfftbib_reformatted, agusample}

%
%
%
%
%

\end{document}